\documentclass{ws-ijmpd}
\usepackage{times}
\usepackage{graphicx}
\usepackage{latexsym}
\usepackage{amsmath}
\usepackage{amsfonts}
\usepackage{amssymb}

\begin{document}

\markboth{L. V. Verozub}
{Hydrodynamic flow as congruence of geodesic lines in Riemannian space-time)}

%
\catchline{}{}{}{}{}
%

\title{HYDRODYNAMIC FLOW AS CONGRUENCE OF GEODESIC LINES IN RIEMANNIAN SPACE-TIME}

\author{L. V. VEROZUB}

\address{\textit{Kharkov National University , 61077 Kharkov, Ukraine \\
Leonid.V.Verozub@univer.kharkov.ua}}

\maketitle

\begin{history}
\received{Day Month Year}
\revised{Day Month Year}
\comby{Managing Editor}
\end{history}

\begin{abstract}
It is shown that smal volume elements  of perfect esentropic fluid  move along geodesic lines of a Riemannian space-time.
\end{abstract}


In papers \cite{Spyrou1,Spyrow2} an interesting idea that fluid motion is equivalent
to geodesic motion of  fluid elements  in non-Euclidean space-time has
been considered. In the non-relativistic  case it follows simply from the fact
that for adiabatic processes in a perfect fluid  with the density $\rho $
and pressure $P$  the first law of classical thermodynamics 
\begin{equation}
d\Pi +Pd\left( \frac{1}{\rho }\right) =0,
\end{equation}%
where $\rho \Pi $  is the fluid specific-energy density, leads to the
equality%
\begin{equation}
\frac{1}{\rho }\boldsymbol{\nabla }P=\boldsymbol{\nabla }\left( \Pi +\frac{P%
}{\rho }\right) .
\end{equation}%
Therefore, along pathes of fluid elements Euler's equation in gravitational
field%
\begin{equation}
\frac{d\mathbf{v}}{dt}=\boldsymbol{\nabla }U-\frac{1}{\rho }\boldsymbol{%
\nabla }P,
\end{equation}%
where $d\mathbf{v/}dt=\partial \mathbf{v}/dt+\mathbf{v\nabla ,}$ takes 
the form%
\begin{equation}
\frac{d\mathbf{v}}{dt}=\boldsymbol{\nabla }U_{eff}
\end{equation}%
where $U_{eff}=U-\left( \Pi +\frac{P}{\rho }\right) $.  Such equations can be
considered as non-relativistic limit of geodesic line of a Riemannian
space-time of curvature other than zero.

However the proof and physical meaning of such fact in relativistic case still remains insufficiently clear.

In this paper we give a simple analysis of the problem and show that small elements  of perfect fluid in adiabatic
processes indeed move along geodesic lines of a Riemannian space-time. 

The concept  of  ``particles''  here means macroscopic small elements of fluid, the motion of which , accordint to the particle-simulation method in hydrodynamics \cite{Monaghan1}, \cite{Monaghan2}, 
 is governed by ordinary differential equations of classical or relativistic dynamics. \footnote{Instead of the traditional continuum assumption, the behavior of a fluid flow can be considered as the motion of a finite mumber of particles uder the influence of interparticles forces which mimic effects of pressure, viscosity, etc. 
Owing to replacement of integration by summation over a number of particles,   continual derivatives  becomes simply  time derivatives along  the particles trajectories. The velocity of the fluid at a given point is the velocity of the particle at this point. The continuity equation is always fulfilled and can consequently be omitted.  Owing to such discratization the motion of particles is governed by means of solutions of ordinary differential equationa of classical or relativistic dynamics. }

Let $d\sigma ^{2}=\eta _{\alpha \beta }dx^{\alpha }dx^{\beta }$ be a
metric differential form of Minkowski space-time $E$ defined in some
differentiable manifold $\mathcal{M\ }$, where $\eta _{\alpha \beta
}=diag(-1,1,1,1)$ is the metric tensor in Cartesian coordinates. Let $w$  be
the  enthalpy per unit volume  and $n$  be particles number density of a perfect
fluid in $E$  so that $w/n$ is the enthalpy per particle. Consider also  in $%
\mathcal{M}$ the second differential metric form 
\begin{equation}
ds^{2}=G_{\alpha \beta }dx^{\alpha }dx^{\beta }  \label{ds2}
\end{equation}%
where $G_{\alpha \beta }=\varkappa ^{2}\eta _{\alpha \beta }$ ,%
\begin{equation}
\varkappa =\frac{w}{nmc^{2}}=1+\frac{\varepsilon }{\rho c^{2}}+\frac{P}{\rho
c^{2}},  \label{xi}
\end{equation}%
$\varepsilon $ is the fluid density energy, $m$ is the mass of the fluid particles,  $c
$ is speed of light, and $\rho=m n$. The form (\ref{ds2}) defines in $\mathcal{M}$ the
structure of a Riemannian space-time $V$ with curvature other than zero.

Let us  show  that the Lagrangian of the motion of particles of perfect fluid 
in adiabatic processes is of the form 
\begin{equation}
L=-mc \left( G_{\alpha \beta }\frac{dx^{\alpha }}{d\lambda }\frac{%
dx^{\beta }}{d\lambda }\right) ^{1/2}d\lambda   \label{Lagrangian_in_V}
\end{equation}%
where $\lambda $ is a parameter along  4-path of particles 

First of all,  it must be noted that the $L$ describes the  motion of the
particles both in $E$ and $V$ . In the first case $G_{\alpha \beta }$ is
some tensor field in $E,$ in the second case it is a fundamental tensor of
the  Riemannian space-time $V.$

In  Minkowski space-time $E$ we can set the parameter $\lambda =\sigma $
which yields the following  Lagrange equations:%
\begin{equation}
\frac{d}{d\sigma }\left( \frac{G_{\alpha \beta }u^{\beta }}{\left( G_{\alpha
\beta }u^{\alpha }u\right) ^{1/2}}\right) -\frac{1}{2\left( G_{\alpha \beta
}u^{\alpha }u\right) ^{1/2}}\frac{\partial G_{\beta \gamma }}{\partial
x^{\alpha }}u^{\beta }u^{\gamma }=0
\end{equation}%
where $u^{\alpha }=dx^{\alpha }/d\sigma $ is 4-velocity of the particles in $%
E.$ Due to the equality $\eta _{\alpha \beta }u^{\alpha }u^{\beta
}=1$ these equations take a simple form%
\begin{equation}
\frac{d}{d\sigma }\left( \varkappa u_{\alpha }\right) -\frac{\partial
\varkappa }{\partial x^{\alpha }}=0
\end{equation}%
where $u_{\alpha }=\eta _{\alpha \beta }u^{\beta }.$ For adiabatic processes 
\cite{Landau} 
\begin{equation}
\frac{\partial }{\partial x^{\alpha}}\left( \frac{w }{n}\right) =\frac{1}{n}\frac{\partial P%
}{\partial x^{\alpha }},
\end{equation}%
and we arrive at the equations of the motion of particles in the form%
\begin{equation}
w\frac{du_{\alpha }}{d\sigma }+u_{\alpha }u^{\beta }\frac{\partial P}{%
\partial x^{\beta }}-\frac{\partial P}{\partial x^{\alpha }}=0.
\label{MotionEquation_in_E}
\end{equation}%
where $du_{\alpha }/d\sigma =\left( \partial u_{\alpha }/\partial
x^{\epsilon }\right) u^{\epsilon }.$ It is the general accepted relativistic
equations of the motion of fluid  \cite{Landau}.

On the other hand,  the Lagrange equations resulting from (\ref{Lagrangian_in_V}%
) are differential  equations of a geodesic in $V.$ If we set $\lambda =s,$
the equations take the standard form of a congruence of geodesic lines :%
\begin{equation}
\frac{du^{\alpha }}{ds}+\Gamma _{\beta \gamma }^{\alpha }u^{\beta }u^{\gamma
}=0,  \label{eqsPart_as_Geodesic}
\end{equation}%
where $du_{\alpha }/ds=\left( \partial u_{\alpha }/\partial x^{\epsilon
}\right) u^{\epsilon }$ and 
\begin{align}
\Gamma _{\beta \gamma }^{\alpha }& =\frac{1}{2}G^{\alpha \epsilon }\left( 
\frac{\partial G_{\epsilon \beta }}{\partial x^{\gamma }}+\frac{\partial
G_{\epsilon \gamma }}{\partial x^{\beta }}-\frac{\partial G_{\beta \gamma }}{%
\partial x^{\epsilon }}\right)  \\
& =\frac{1}{\varkappa }\left( \frac{\partial \varkappa }{\partial x^{\gamma }%
}\delta _{\beta }^{\alpha }+\frac{\partial \varkappa }{\partial x^{\beta }}%
\delta _{\gamma }^{\alpha }-\eta ^{\alpha \epsilon }\frac{\partial \varkappa 
}{\partial x^{\epsilon }}\eta _{\beta \gamma }\right) .
\end{align}%

The component 
\begin{equation}
\Gamma _{00}^{1}=-\frac{1}{\rho c^{2} }\frac{\partial P}{\partial x^{1}}
\end{equation}%
and  eqs. (\ref{eqsPart_as_Geodesic}) lead  to standard Euler equation  in
non-relativistic limit.

It ie easy to verify that a covariant derivative of tensors $G_{\alpha\beta}$ is equal to zero identically.

Thus, the motion of fluid particles in pseudo-Euclidean space-time are at
the same time equations of the motion of these particles along geodesic lines
of the Riemannian space-time $V$. \footnote{The reviewer of this paper has kindly specified to me that some conformal factor, namely ``density+interaction energy density'',  for the space-time metric in an elastic medium was used in DeWitt's paper \cite{DeWitt}, and later by Brown and Marolf \cite{Brown} in the  examination of the moving of material medium as a frame of reference in general relativity.}

It is easily to expand the above result to the case of the motion of fluid
in arbitrary Riemannian space-time with a metric tensor $g_{\alpha \beta
}(x).$ In this case 
\begin{equation}
G_{\alpha \beta }=\varkappa ^{2}g_{\alpha \beta }.  \label{GalhabetaGeneral}
\end{equation}%
If the parameter $\lambda =\sigma ,$ the Lagrange equations (\ref{Lagrangian_in_V}) yield 
the equation which differ from (\ref{MotionEquation_in_E} ) only by 
covariant derivatives instead the ordinary
ones. These equations together with continuity equation are standard
equations of fluid motion in General relativity.

 If we set $\lambda =s,$ we
will obtain these equations in the form of standard geodesic equations of
the Riemannian space-time with the metric tensor $G_{\alpha \beta }$.  More
suitable to us as a parameter  is the coordinate time $x^{0}$ instead $s$ so that the
equations take the form%
\begin{equation}
\overset{..}{x}^{\alpha }+\left( \Gamma _{\beta \gamma }^{\alpha
}-c^{-1}\Gamma _{\beta \gamma }^{0}\overset{.}{x}^{\alpha }\right) \overset{.%
}{x}^{\beta }\overset{.}{x}^{\gamma }=0  \label{geodesicEquationInV}
\end{equation}%
where $\overset{.}{x}^{\alpha }=dx^{\alpha }/dt$.  Zero component of these
equations is satisfied identically, and rest equations are the ones for the
3-spacial velocity.

To make sure that these geodesic equations together with the continuity
equations give the generally accepted description of fluid  in General
Relativity we consider the conditions of a fluid equilibrium in a given
spherically-symmetric gravitational field. Since $\overset{.}{x}^{\alpha }=%
\overset{.}{x}^{\alpha }=0,$ the conditions of the equilibrium in spherical
coordinates are%
\begin{equation}
\Gamma _{00}^{1}=\frac{1}{2}G^{11}\left( G_{00}\right) ^{\prime }=0
\end{equation}%
where the prime denotes a derivative with respect to the radial distance $r.$
It means that $\left( \varkappa ^{2}g_{00}\right) ^{\prime }=0,$ or%
\begin{equation}
\frac{\varkappa ^{\prime }}{\varkappa }=-\frac{\left( g_{00}\right) ^{\prime
}}{2\ g_{00}}.
\end{equation}%
Due to the equality  $\varkappa ^{\prime }/\varkappa =P^{\prime }/\rho c^{2}
\varkappa .$ and taking into account (\ref{xi}) we obtain for gas with $%
\varepsilon =0$ the standard equation of the equilibrium in General
Relativity \cite{Weinberg}%
\begin{equation}
\frac{\left( g_{00}\right) ^{\prime }}{\ g_{00}}=-\frac{2P^{\prime }}{\rho
c^{2}+P}.
\end{equation}

Thus, the motion of an ideal fluid particles in a space-time with the metric tensor $g_{\alpha\beta}(x)  $ at adiabatic processes take place
along geodesic lines of the Riemannian space-time with the metric tensor (%
\ref{GalhabetaGeneral}).

This fact is of fundamental importance since it shows a bimetric nature of
motion particles under the influence of force fields.
Besides that, it    gives new insight into a description of
hydrodynamic flow, and allows to solve many tasks more  easily.

As a simple example consider a spherically-symmetric 
accretion a gas onto super-massive compact object in General Relativity. In the
spherical coordinates $(t,r,\theta ,\varphi )$ the Lagrangian (\ref{Lagrangian_in_V}) is of the form%
\begin{equation}
L=-mc\ \varkappa ^{2}\left[ A\overset{.}{r}^{2}+r^{2}(\overset{.}{\theta }%
^{2}+\sin ^{2}\theta \overset{.}{\varphi }^{2})-c^{2}C\right] ^{1/2}
\end{equation}%
where $C=1-r_{g}/r$ ,  $A=C^{-1},$ $r_{g}=2\gamma M/c^{2},$ $\gamma $ is the
gravitational constant,   $M$ is the mass of the central object. Here  $%
\varkappa $ is the function of the radial coordinate $r$, and the points
denote the derivative with respect to time $t.$ Owing to conservation of
the energy $E$ and angular momentum $J$ of gas particles at the geodesic
(free) motion in $V$, the equations of the motion in the plane $\theta =\pi
/2$ takes the form 
\begin{align}
\overset{.}{r}^{2}& =c^{2}C^{2}\left[ 1-\left(\varkappa ^{2}C/\overline{E}%
^{2}\right)   \left(1+\overline{J}^{2}/\varkappa ^{2}\overline{r}^{2}\right)\right] 
\label{MotionEquations} \\
\overset{.}{\varphi }& =cC\overline{J} r_{g}/\overline{r}^{2}\overset{\_%
}{E}  \notag
\end{align}%

\begin{figure}[tbh]
\centering
 \includegraphics[width=8cm,height=6cm]{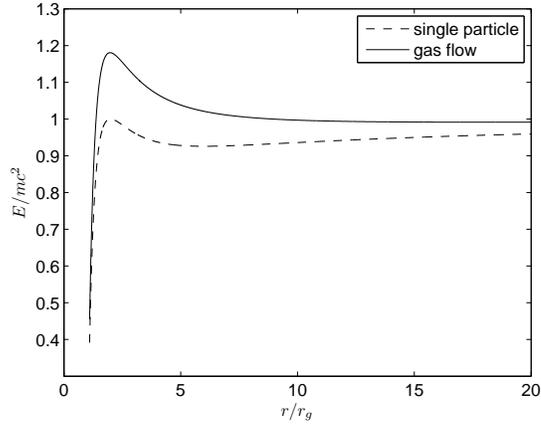}
\caption{ An effective
potential for the motion of  particles of gas flow  for Schwarzschild
solution as compared with the efective potential of single particles. The
particles energy $\overline{E}=1$,  the angular momentum $\protect\overline{J}$ $=2$.}
\end{figure}

where $\overline{E}=E/mc$, $\overline{r}=r/r_{g}$
$\overline{J}=J/r_{g}mc.$ These equations
give more information about gas motion than Bernoulli equation, and allow
to see easily the difference between motion of single particles and motion
of a flow . In particular,  if  we set  in (\ref{MotionEquations}) $\overset{.}{r}=0$ 
we obtain the effective potential \cite{Novikov} for the motion of the flow particles, i.e. $\overline{E}^{2},$
as a function of $\overline{r}$ : 
\begin{equation*}
\overline{E}^{2}\mathcal{(}\overline{r})=\varkappa ^{2}(1-1/\overline{r})(1+%
\overline{J}^{2}/\varkappa ^{2}\overline{r}^{2}).
\end{equation*}%

Fig. 1 shows the effective potential for a single particle and
for the relativistic  gas flow  where the function $\varkappa =1/3\overline{r}$. It
shows that the function $\varkappa (r)$ strongly influences  on the postion
of the minimum of the effective potential that defines the position of
stable circular orbits of the gas around of the central object. 
The minimum of secular orbits of the gas lies at the distance  $\overline{r}=18.7$,  while
the one for the single particles  lies at the distance  $\protect\overline{r}$ $=$ $6$.


\end{document}